\documentclass[aps,prb,twocolumn,amsmath,superscriptaddress]{revtex4}
\usepackage{graphicx}
\usepackage{color}
\usepackage{epstopdf}
\usepackage{xcolor}
\usepackage{amsmath}

\begin{document}

\title{Resonant photoluminescence and dynamics of a hybrid Mn-hole spin in a positively charged magnetic quantum dot}

\author{A. Lafuente-Sampietro}
\affiliation{CNRS, Institut N\'eel, F-38042 Grenoble, France.}
\affiliation{Universit\'{e} Grenoble Alpes, Institut N\'eel, F-38042 Grenoble, France}

\author{H. Boukari}
\affiliation{CNRS, Institut N\'eel, F-38042 Grenoble, France.}
\affiliation{Universit\'{e} Grenoble Alpes, Institut N\'eel, F-38042 Grenoble, France}

\author{L. Besombes}
\email{lucien.besombes@grenoble.cnrs.fr}
\affiliation{CNRS, Institut N\'eel, F-38042 Grenoble, France.}
\affiliation{Universit\'{e} Grenoble Alpes, Institut N\'eel, F-38042 Grenoble, France}
\date{\today}

\begin{abstract}

We analyze, through resonant photoluminescence, the spin dynamics of an individual magnetic atom (Mn) coupled to a hole in a semiconductor quantum dot. The hybrid Mn-hole spin and the positively charged exciton in a CdTe/ZnTe quantum dot forms an ensemble of $\Lambda$ systems which can be addressed optically. Auto-correlation of the resonant photoluminescence and resonant optical pumping experiments are used to study the spin relaxation channels in this multilevel spin system. We identified for the hybrid Mn-hole spin an efficient relaxation channel driven by the interplay of the Mn-hole exchange interaction and the coupling to acoustic phonons. We also show that the optical $\Lambda$ systems are connected through inefficient spin-flips than can be enhanced under weak transverse magnetic field. The dynamics of the resonant photoluminescence in a p-doped magnetic quantum dot is well described by a complete rate equation model. Our results suggest that long lived hybrid Mn-hole spin could be obtained in quantum dot systems with large heavy-hole/light-hole splitting.

\end{abstract}

\maketitle

\section{Introduction}

Individual localized spins are promising for the implementation of emerging quantum information technologies in the solid state \cite{Petta2005,Veldhorst2015,Saeedi2013,Bar-Gill2013}. In this field, magnetic dopants in conventional semiconductors present many desirable features, such as reproducible quantum properties, stability, and potential scalability for further applications \cite{Koenrad2011}. Thanks to their point-like character, a longer spin coherence time (compared to carriers’ spins) can also be expected at low temperature, making them potentially good systems to store quantum information. Semiconductor quantum dots (QDs) permit efficient electrical or optical manipulation of individual carrier spins \cite{Atature2006,Press2008,Gerardot2008,DeGreve2011}. It has been shown that the optical properties of a QD can also be used to control the spin state of individual \cite{Besombes2004,LeGall2011,Kudelski2007,Krebs2009,Baudin2011,Kobak2014} or pairs \cite{Besombes2012,Krebs2013} of magnetic atoms. The spin of a magnetic atom in a QD can be prepared by the injection of spin polarized carriers and its state can be read through the energy and polarization of the photons emitted by the QD \cite{Besombes2015,Reiter2013,Goryca2009}. The insertion of a magnetic atom in a QD where the strain or the charge states can be controlled also offers degrees of freedom to tune the properties of the localized spin such as its magnetic anisotropy responsible for the spin memory at zero magnetic field \cite{Oberg2014}.

Positively charged magnetic QDs present a large magnetic anisotropy induced by the exchange interaction between the confined heavy-hole and the spin localized on a magnetic atom \cite{Leger2005,Vyborny2012}. It has been shown in charge tunable Mn-doped CdTe/ZnTe QDs that the hybrid Mn-hole spin can be prepared by optical orientation under excitation with circularly polarized light \cite{Varghese2014}. It was also demonstrated that a positively charged Mn-doped QD forms an ensemble of optical $\Lambda$ systems that can be independently addressed \cite{Lafuente2015}. This level structure suggests that the hybrid Mn-hole spin states could be coherently manipulated using two resonant optical fields \cite{Houel2014}. For an efficient coherent control and a practical use of this hybrid spin in a quantum device one would require however sufficiently long relaxation and coherence times for the two ground states of the $\Lambda$ systems.

In this article, resonant photoluminescence (PL) of the positively charged exciton (X$^+$) coupled to a single Mn atom in charge tunable CdTe/ZnTe QDs \cite{Varghese2014} is exploited to analyze the spin relaxation mechanisms of the coupled hole and Mn spins. Auto-correlation of the resonant PL and resonant optical pumping experiments reveal an efficient spin relaxation channel of the Mn-hole system: a Mn-hole flip-flop induced by the interplay of the Mn-hole exchange interaction and the lattice deformation induced by acoustic phonons. A model predicts Mn-hole flip-flops in the nanosecond range, in agreement with the spin dynamics observed experimentally. We also show that the optical $\Lambda$ systems are connected by inefficient forbidden spin-flips which are enhanced under a weak transverse magnetic field.

The article is organized as follows: The sample and experiments are presented in Sec. II. The spin structure of the positively charged Mn-doped QDs is shortly described in Sec. III. The dynamics of the resonant PL probed with auto-correlation measurements is described in Sec. IV. In Sec. V we discuss resonant optical pumping experiments allowing probing the initialization and relaxation of coupled hole and Mn spins. In Sec. VI we present a model describing acoustic phonon induced Mn-hole flip-flops and show that this mechanism is responsible for the strong PL observed under resonant excitation of X$^+$-Mn. Finally, in Sec. VII we model the spin dynamics of positively charged Mn-doped QDs under resonant excitation and compare the calculated autocorrelation response and optical pumping signal with experimental results.

\section{Sample and experiments}

\begin{figure}[hbt]
\includegraphics[width=3.2 in]{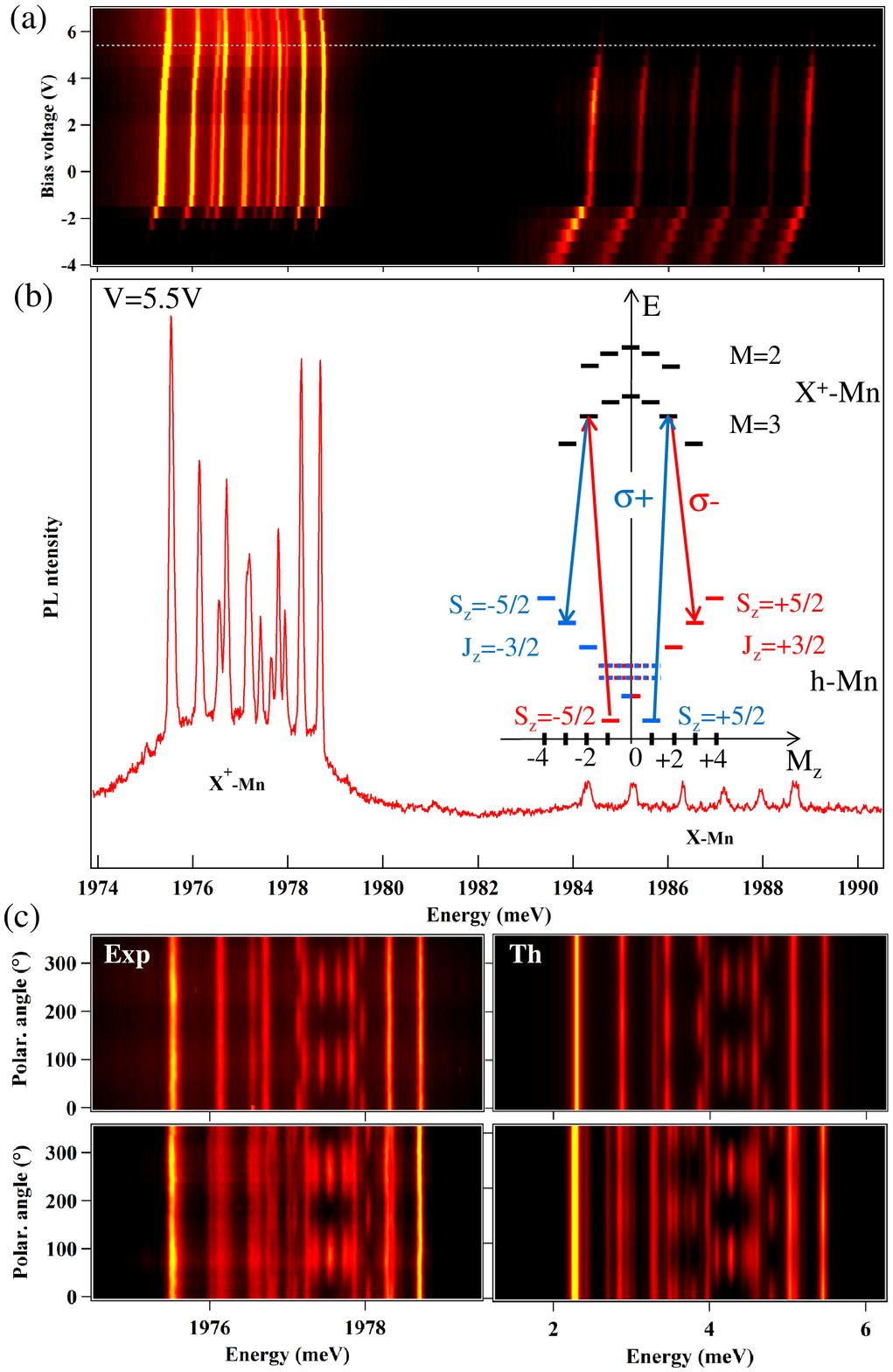}
\caption{(a) Color scale plot of the PL intensity of the studied Mn doped QD inserted in Schottky structure showing the emission of the neutral ($X-Mn$) and positively charged ($X^+-Mn$) exciton as a function of energy and bias voltage. (b) PL of the Mn-doped QD under a positive bias voltage of V=5.5V. Inset: Scheme of the energy levels of the ground ($h-Mn$) and excited states ($X^+-Mn$) in a positively charged Mn-doped QD as a function of their angular momentum ($M_z$). (c) Experimental (left) and calculated (right) color-scale plot of the linear polarization dependence of the PL of $X^+-Mn$ at B=0T (top) and under a transverse magnetic field B$_\perp$=0.42T (bottom), maximum transverse field used in the spin dynamics experiments. The parameters used in the calculation are listed in table \ref{paraQD}.}
\label{Fig1}
\end{figure}

The sample consists of Mn-doped self-assembled CdTe QDs grown by molecular beam epitaxy on a p-doped ZnTe (001) substrate according to the procedure described in ref. 27. A bias voltage is applied between a 5 nm gold Schottky gate deposited on the surface of the sample and the p-doped substrate to control the charge state of the QDs. Under a positive bias, as we already reported in ref. 24, a single hole is trapped in the magnetic QDs and only the emission of the positively charged exciton is observed (figure~\ref{Fig1}(a)).

Individual QDs containing one Mn are isolated using micro-spectroscopy techniques \cite{LeGall2010}. A high refractive index hemispherical solid immersion lens is mounted on the surface of the sample to enhance the spatial resolution and the collection efficiency of single dot emission. The QDs are excited with a tunable continuous wave ($cw$) laser tuned to an excited state of the dots or on resonance with the ground state of the positively charged exciton. The resulting collected PL is dispersed and filtered by a one-meter double monochromator.

The temporal statistics of the emission of photons is analyzed through photon-correlation measurements using a Hanbury Brown and Twiss (HBT) setup with a resolution of about 0.8 ns. Under our experimental conditions, with photon counts rates of a few kHz, the measured photon pair time distribution yields, after normalization, to the second order correlation function of the PL intensity, $g^{(2)}(\tau)$.

For optical pumping experiments, trains of resonant circularly polarized light are prepared with electro-optic or acousto-optic modulators with a switching time of about 10 ns and the resonant PL is detected with a fast avalanche photodiode. Permanent magnet mounted on a translation stage are used to apply a weak magnetic field in Faraday or Voight configurations.

\section{Spin structure of a positively charged Mn-doped quantum dot}

A Mn atom in a strained self-assembled CdTe QD exhibits a fine structure dominated by a weak magnetic anisotropy with an easy axis along the QD axis \cite{LeGall2010,LeGall2009,Goryca2014}. Neglecting the tetrahedral crystal field of the CdTe matrix \cite{Qazzaz1995,Causa1980}, this fine structure is described by the effective spin Hamiltonian

\begin{eqnarray}
\label{cf}
{\cal H}_{Mn,CF}=D_0S^2_z+E(S_y^2-S_x^2)
\end{eqnarray}

\noindent Here, $D_0$ depicts the effect of the biaxial strain and $E$ describes the anisotropy of the strain in the plane of the QD. $D_0$ varies from 0 $\mu$eV (strain free QD \cite{Besombes2014}) to 12 $\mu$eV (strained CdTe layer matched on a ZnTe substrate \cite{LeGall2009}) and typical values around 7 $\mu$eV are usually observed in CdTe/ZnTe QDs \cite{Jamet2013,Goryca2014}.

When a hole is trapped in a QD containing a single Mn, the spin structure is controlled by the Mn-hole exchange interaction that reads

\begin{eqnarray}
{\cal H}_{hMn}^{ex}=I_{hMn}\vec{S}\cdot\vec{J}
\end{eqnarray}

\noindent with I$_{hMn}$ the exchange energy between the hole and the Mn ($S=5/2$) and $\vec{J}$ the hole spin operator. In the presence of heavy-hole/light-hole mixing, $\vec{J}$, represented in the basis of the two low energy heavy-hole states, is related to the Pauli matrices by $J_z=3/2\tau_z$ and $J_{\pm}= \xi \tau_{\pm}$ with $\xi=-2\sqrt{3}e^{-2i\theta}\rho_c/\Delta_{lh}$. $\rho_c$ is the coupling energy between heavy holes and light holes separated by an effective energy splitting $\Delta_{lh}$. $\theta$ is the angle relative to the [110] axis of the principal axis of the anisotropy (shape and/or strain) responsible for the heavy-hole/light-hole mixing \cite{Fernandez2006,Leger2007}. For a weak valence band mixing, the Mn-hole energy levels are mainly controlled by $I_{hMn}S_zJ_z$ and form a spin ladder with a quantization axis along the QDs growth direction. These states are labelled $|S_z,J_z\rangle$.

The PL of a positively charged Mn-doped QD is presented in figure~\ref{Fig1}(b). When an exciton is injected in the QD loaded with a single hole, one has to consider the X$^+$-Mn complex. The energy levels of such a complex, where the two hole spins are paired in an antiparallel configuration, is dominated by the electron-Mn exchange interaction

\begin{eqnarray}
{\cal H}_{eMn}^{ex}=I_{eMn}\vec{S}\cdot\vec{\sigma}
\end{eqnarray}

\noindent with $\vec{\sigma}$ the electron spin and $I_{eMn}$ the exchange energy between the electron and the Mn spins. The twelve e-Mn states are split into a ground state sextuplet (total spin M=3) and a fivefold degenerated manifold (total spin M=2) (see inset of Fig.~\ref{Fig1}(b)). These energy levels are labelled $|M,M_z\rangle$.

\begin{table}[t] \centering
\caption{Values of the parameters used to model the positively charged Mn-doped QD presented in figure \ref{Fig1}. I$_{eMn}$, I$_{hMn}$, $\frac{\rho_c}{\Delta_{lh}}$, $\theta$, $\eta$ and $T_{eff}$ are chosen to reproduce the linear polarization intensity map of Fig.~\ref{Fig1}(c). The other parameters cannot be extracted from the PL measurements and values for typical Mn-doped QDs are chosen for the calculation of the spin dynamics presented in section VII.}
\renewcommand{\arraystretch}{1.0}

\begin{tabular}{lcr}
\hline\hline
Exchange interaction & I$_{eMn}$ &  -175$\mu eV$  \\
& I$_{hMn}$&  345$\mu eV$  \\
Valence band mixing & $\frac{\rho_c}{\Delta_{lh}}$ &  0.09  \\
& $\theta$ &  0 $^{\circ}$  \\
Hole perturbation & $\eta$ &  30$\mu eV$   \\
Effective temperature & $T_{eff}$ &  20 K \\
\hline
g factors  & $g_{e}$ &  -0.4  \\
& $g_{h}$ &  -0.6  \\
& $g_{Mn}$  &  2 \\
Mn fine structure & $D_0$  &  7 $\mu eV$ \\
& $E$  &  1.5 $\mu eV$  \\
\hline\hline
\end{tabular}
\label{paraQD}
\end{table}

The exchange coupling between the holes and the Mn introduce also a perturbation of their wave functions \cite{Besombes2005,Trojnar2013,Besombes2014}, that can be represented for one hole by an effective spin Hamiltonian ${\cal H}_{scat}=-\eta S_z^2$ with $\eta>$0. This perturbation has to be taken into account twice for X$^+$-Mn where two holes interact with the Mn. This perturbation affects the energy of the optical recombination of X$^+$-Mn to the Mn-hole ground state and can be observed in the emission spectra \cite{Besombes2014}.

Values of $I_{hMn}$, $I_{eMn}$, $\rho_c/\Delta_{lh}$ and $\eta$ for a given QD can be obtained by comparing the linear polarization dependence of the experimental PL data to the optical transition probabilities calculated with the discussed effective spin model (Fig.~\ref{Fig1}(c)) \cite{Varghese2014}. A Boltzmann distribution function $P^i_{eMn}=e^{-E^i_{eMn}/k_BT_{eff}}/\sum_{i}e^{-E^i_{eMn}/k_BT_{eff}}$ with an effective spin temperature $T_{eff}$ is used to describe the population of the emitting states (electron-Mn energy levels $E^i_{eMn}$). The obtained parameters are listed in table \ref{paraQD} for the QD presented in figure~\ref{Fig1}.

\section{Dynamics of the resonant photoluminescence of X$^+$-Mn.}

\begin{figure}[hbt]
\includegraphics[width=3.2 in]{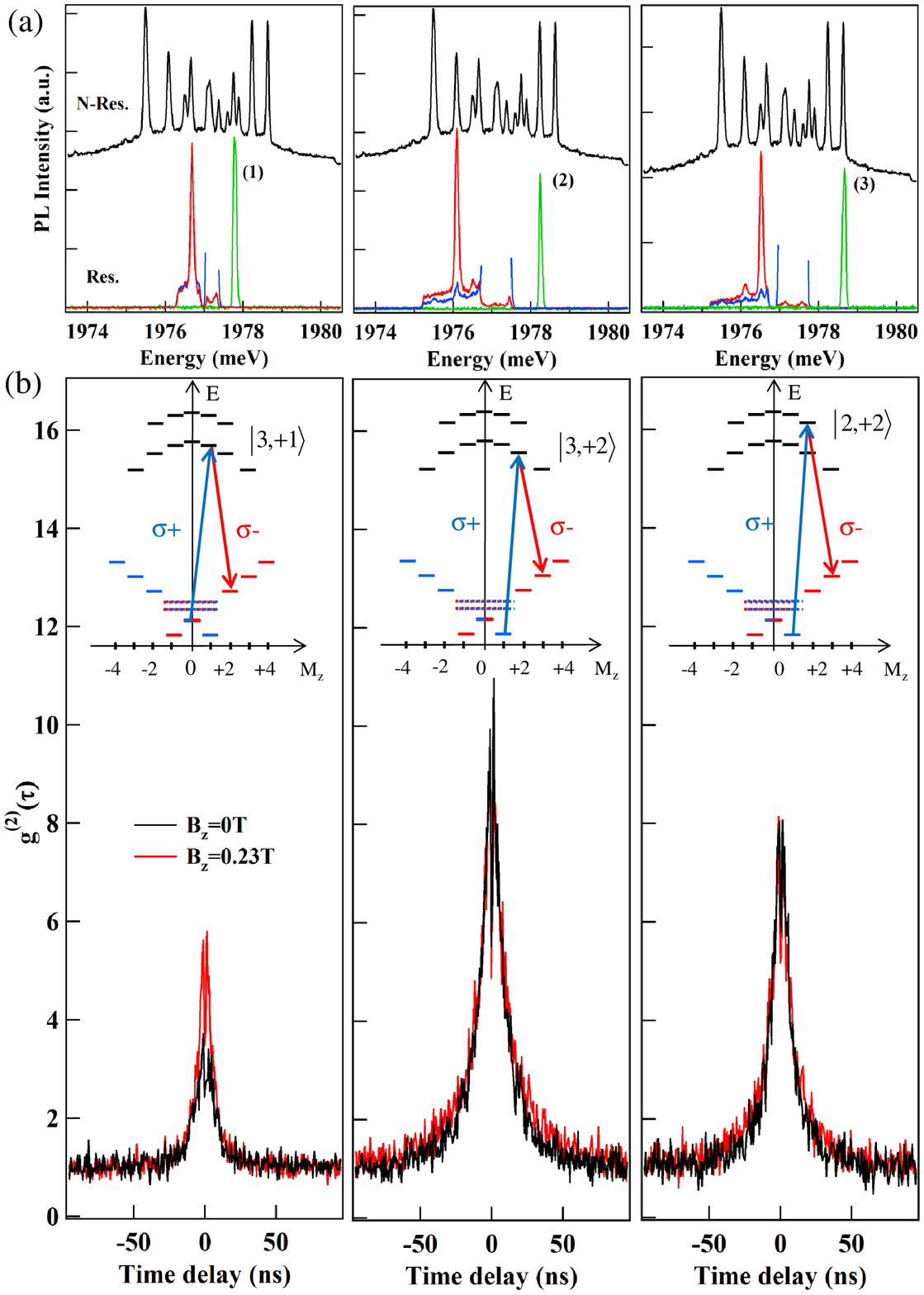}
\caption{(a) Non resonant (N-Res.) and resonant (Res.) PL of X$^+$-Mn. Co (blue) and cross (red) circularly polarized PL spectra are collected for three different energies of the CW resonant laser (green). (b) Auto-correlation of the resonant PL for a cross circularly polarized excitation and detection of the electron-Mn states $|3,+1\rangle$, $|3,+2\rangle$ and $|2,+2\rangle$. Insets: Energy levels of X$^+$-Mn and identification of the three resonances observed in (a) corresponding to the optical $\Lambda$ systems associated with the states $|3,+1\rangle$, $|3,+2\rangle$ and $|2,+2\rangle$.}
\label{Fig2}
\end{figure}

To access the dynamics of the hybrid Mn-hole spin we exploit the resonant PL of X$^+$-Mn. When scanning a resonant laser through the high energy transitions of the X$^+$-Mn, three absorption resonances are observed \cite{Lafuente2015}. These, labelled (1), (2) and (3) in Fig.~\ref{Fig2}(a), give rise to a strong resonant PL which is cross circularly polarized with the excitation, except for an excitation on (1) which produces unpolarized PL. The corresponding energy levels involved in these absorption are identified in Fig.~\ref{Fig2}(b). They correspond, for a $\sigma+$ laser, to the successive resonant excitation of the electron-Mn levels $|3,+1\rangle$, $|3,+2\rangle$ and $|2,+2\rangle$ \cite{Lafuente2015}. These states can be expressed as linear combinations of the Mn and electron spins $|S_z,\sigma_z\rangle$ coupled by a flip-flop:

\begin{eqnarray}
|3,+1\rangle&=& \frac{1}{\sqrt{6}}(\sqrt{4}|+1/2,\uparrow_e\rangle+\sqrt{2}|+3/2,\downarrow_e\rangle)\\
|3,+2\rangle&=& \frac{1}{\sqrt{6}}(\sqrt{5}|+3/2,\uparrow_e\rangle+\sqrt{1}|+5/2,\downarrow_e\rangle)\\
|2,+2\rangle&=& \frac{1}{\sqrt{6}}(\sqrt{1}|+3/2,\uparrow_e\rangle-\sqrt{5}|+5/2,\downarrow_e\rangle)
\end{eqnarray}

Each of these electron-Mn states are connected with circularly polarized optical transitions to two Mn-hole states in the ground state of the QD. For instance $|3,+2\rangle$ is connected to $|+\frac{3}{2},\Uparrow_h\rangle$ with $\sigma-$ photons and to $|+\frac{5}{2},\Downarrow_h\rangle$ with $\sigma+$ photons. This level structure forms an optical $\Lambda$ system.

Under resonant excitation of one high energy level of X$^+$-Mn, only one cross-circularly polarized emission line is observed. It corresponds to the optically allowed recombination on the second branch of the $\Lambda$ system. This recombination occurs with a flip-flop of the electron and Mn spins \cite{Varghese2014,Lafuente2015}. The energy splitting between the resonant absorption and the emission corresponds to the splitting between the two ground states of the $\Lambda$ system. It is given by 4$\times$3/2$I_{hMn}$($\approx$2.1 meV for the studied QD) for an excitation of $|3,+2\rangle$ or $|2,+2\rangle$ and 2$\times$3/2$I_{hMn}$($\approx$1.05 meV for the studied QD) for an excitation of $|3,+1\rangle$. For an excitation of $|3,+2\rangle$ or $|2,+2\rangle$, the weak co-polarized PL signal, which depends on the excitation intensity, comes from a possible direct excitation of the low energy branch of the $\Lambda$ system through the acoustic phonon side-band \cite{Besombes2001}.

For an isolated $\Lambda$ system, under resonant excitation of one of the branch, a fast optical pumping controlled by the generation rate and the radiative lifetime of the excited state is expected: The population is expected to be stored in the level which is not excited and the resonant PL should vanish. In the case of X$^+$-Mn, the PL intensity observed under resonant excitation of the high energy branch of the $\Lambda$ systems is similar to the PL intensity obtained under non-resonant excitation. This suggests a very inefficient optical pumping of the Mn-hole spin and an efficient spin-flip mechanism which links the two ground states of the $\Lambda$ systems.

\begin{figure}[hbt]
\includegraphics[width=3.1 in]{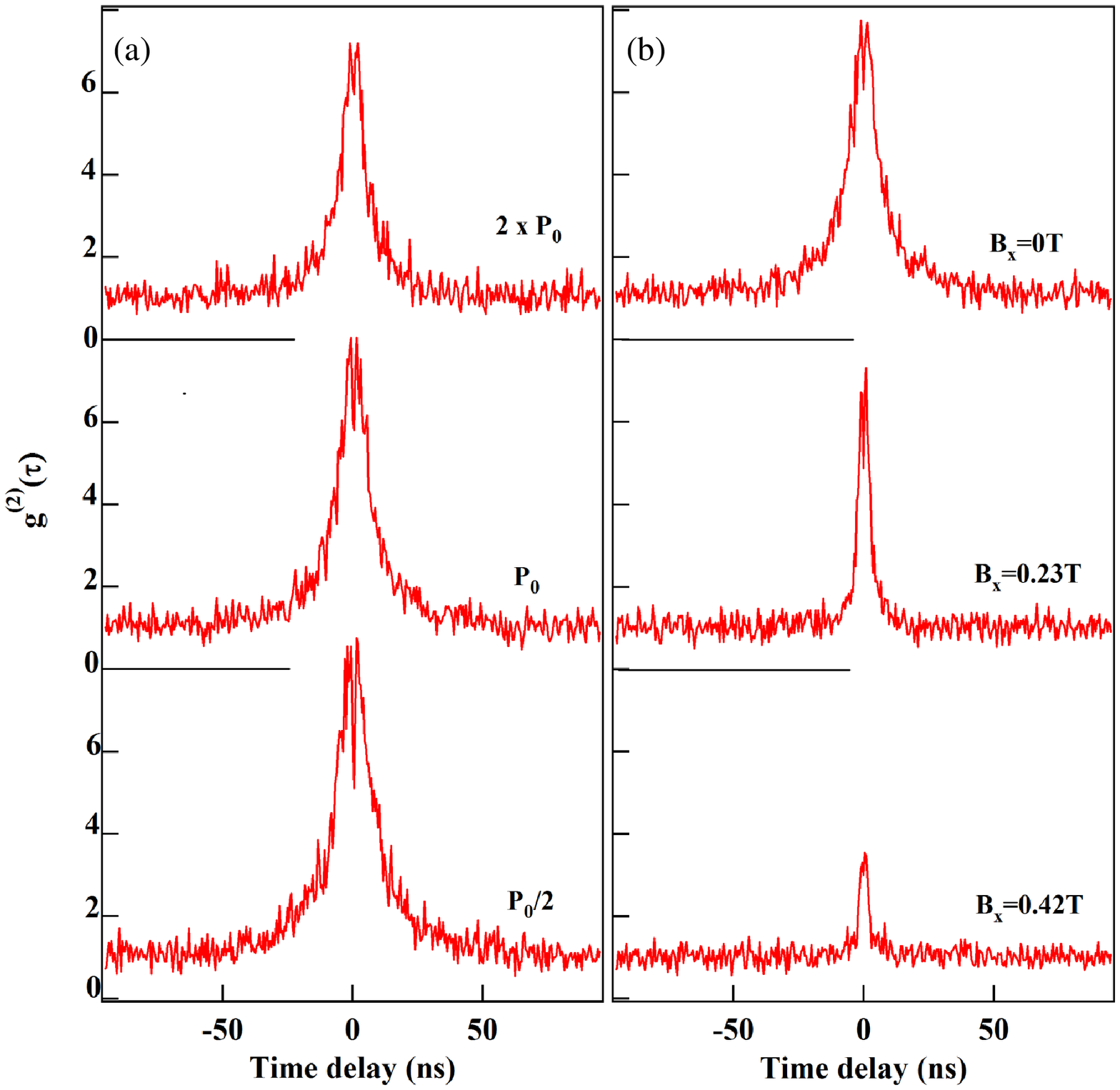}
\caption{Excitation power dependence (a) and transverse magnetic field dependence (b) of the auto-correlation of the resonant PL obtained for an excitation on the high energy branch of the $\Lambda$ level system associated to the electron-Mn state $|2,+2\rangle$.}
\label{Fig3}
\end{figure}

The dynamics of the Mn spin coupled to carriers was first analyzed, under resonant optical excitation, through the statistics of the time arrival of the photons given by the second order correlation function of the resonant PL intensity, $g^{(2)}(\tau)$. For the three resonant excitation conditions reported in Fig.\ref{Fig2}(b), $g^{(2)}(\tau)$ is mainly characterized by a large photon bunching with a full width at half maximum (FWHM) in the 20 ns range. The amplitude of the bunching reaches 9 for line (2) and is slightly weaker for the two other lines. This large bunching, reflecting an intermittency in the emission of the QD, is not sensitive to a longitudinal magnetic field B$_z$ except for an excitation on (1).

The presence of a photon bunching is at first sight surprising: under resonant excitation of an isolated $\Lambda$ system, an anti-bunching of the resonant PL controlled by the transfer time between the two ground states is indeed expected. For X$^+$-Mn, the observed short anti-bunching (dip near zero delay in the autocorrelation function observed at low excitation power in figure~\ref{Fig3}(a)) suggests a fast transfer time in the nanosecond range between the two ground states of the $\Lambda$ systems.

In the presence of a transfer process connecting the two Mn-hole ground states in a nanosecond time-scale, the photon bunching can be explained by leaks outside the resonantly excited $\Lambda$ system. Under $cw$ excitation, the population is cycled inside the $\Lambda$ system until a spin flip occurs and drives the carrier-Mn spin out of the $\Lambda$ levels under investigation. The resonant PL is then switched off until multiple spin-flips drives back the carriers and Mn spin inside the $\Lambda$ system under excitation. The selected QD line can be either in a ON or OFF state depending on the fluctuations of the carrier and Mn spins. The amplitude of the bunching is then given by $\Gamma_{Out}/\Gamma_{In}$ the ratio of the transition rates from OFF to ON ($\Gamma_{In}$) and from ON to OFF ($\Gamma_{Out}$). An amplitude of bunching larger than 1 is expected for the multilevel system considered here where, after a spin relaxation, multiple spin flips are in average required to come back to the initial state ($\Gamma_{In}<\Gamma_{Out}$). Within this picture, the width of the bunching is a measurement of the escape time out of the considered $\Lambda$ level system.

A weak transverse magnetic field, $B_x$, significantly reduces the width of the bunching signal (Fig.\ref{Fig3}(b)). As the spin of the Mn-hole complex is highly anisotropic, with a large energy splitting induced by the exchange interaction $I_{hMn}S_z.J_z$, the weak transverse magnetic field mainly affects the electron-Mn dynamics in the excited state of the charged QD. Indeed, the transverse magnetic field couples the different electron-Mn states and induces a leak outside the resonantly excited $\Lambda$ system. Both spin-flips within the Mn-hole (ground state) and the electron-Mn (excited state) systems can contribute to the bunching signal. The significant effect of the weak transverse field shows that the probability of presence in the excited state of the $\Lambda$ system is large. This is consistent with the large excitation intensity used for these auto-correlation measurements which require a high photon count rate.

A slight reduction of the width of the bunching signal is also observed with the increase of the excitation power (Fig.\ref{Fig3}(a)). This shows that the leaks outside a given $\Lambda$ system slightly increases with the probability of presence of the positively charged exciton in the QD.

\section{Resonant optical pumping and relaxation of the hybrid Mn-hole spin}

\begin{figure}[hbt]
\includegraphics[width=3.4in]{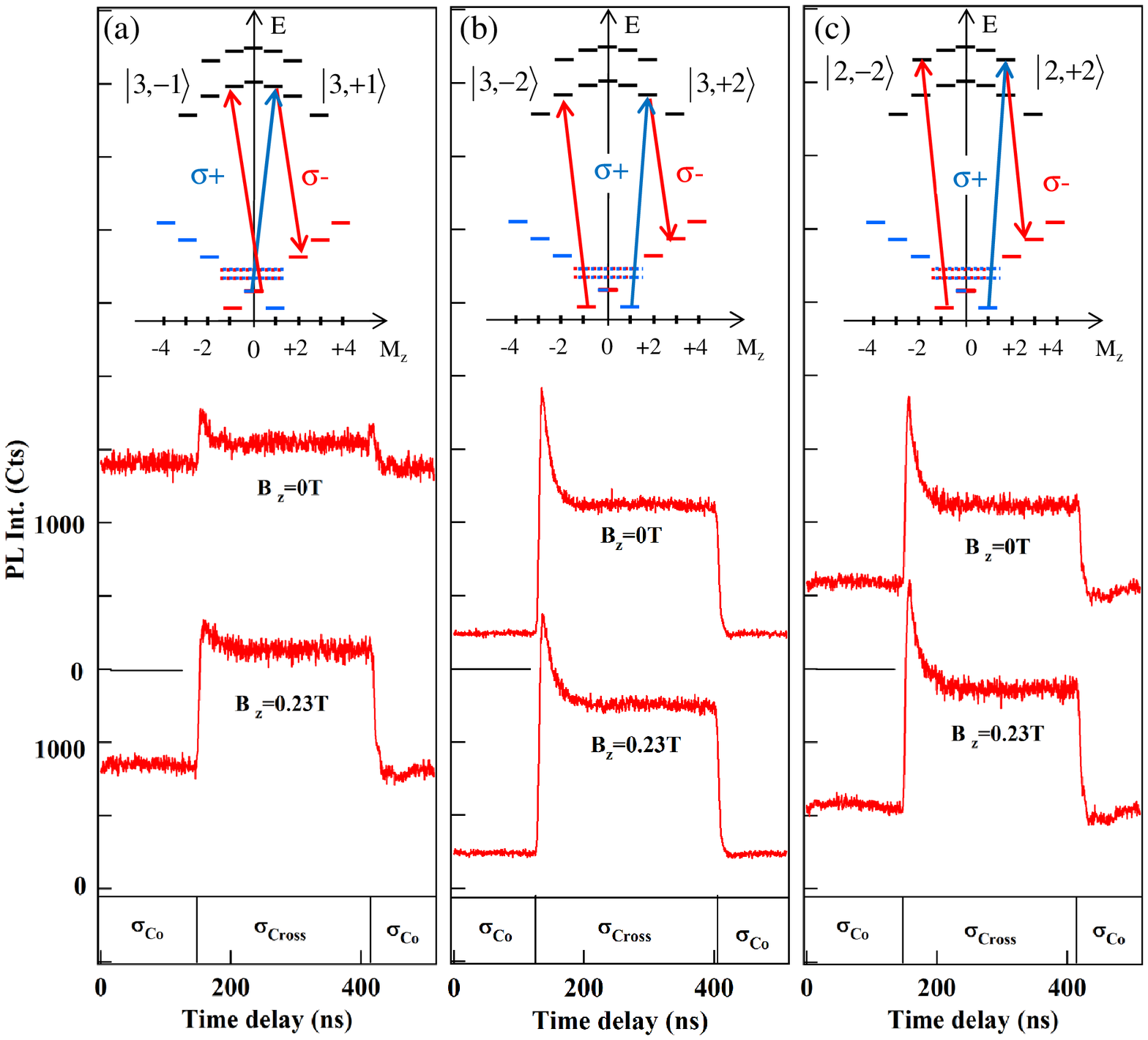}
\caption{Resonant optical pumping transients obtained under circular polarization switching of the resonant excitation for the three $\Lambda$ systems identified in figure \ref{Fig2} at zero field and under a weak longitudinal magnetic field B$_z$=0.23T. The insets present the corresponding states which are resonantly excited and detected in $\sigma-$ polarization.}
\label{Fig4}
\end{figure}

Resonant optical pumping experiments were done to estimate how long it takes, after a spin-flip, to the hybrid Mn-hole spin to relax back inside the resonantly excited $\Lambda$ system. A demonstration of resonant optical pumping of the Mn-hole system was first done by exciting the high energy branch of the $\Lambda$ systems with trains of resonant light, alternating the circular polarization and recording the circularly polarized PL of the low energy branch. As observed in Fig.~\ref{Fig4}, for an excitation on resonance with the electron-Mn states $|3,+2\rangle$ or $|2,+2\rangle$, switching the polarization of the excitation from co to cross circular produces a change of the PL intensity with two transients: first, an abrupt PL increase (or decrease), reflecting the population change of the observed spin-polarized charged excitons; then a slower transient with a characteristic time of a few tens of nanoseconds, depending on the laser excitation power.

The progressive decrease of the resonant PL intensity is the signature of an optical pumping of the Mn-hole spin: the Mn-hole state which is optically addressed is partially emptied when the population is ejected out of the excited $\Lambda$ system. As presented in figure~\ref{Fig4}, this pumping signal is not sensitive to a longitudinal magnetic field B$_z$ except for an excitation of $|3,\pm1\rangle$ where a significant intensity difference between co and cross circular polarization is only observed under a weak B$_z$.

\begin{figure}[hbt]
\includegraphics[width=3.3in]{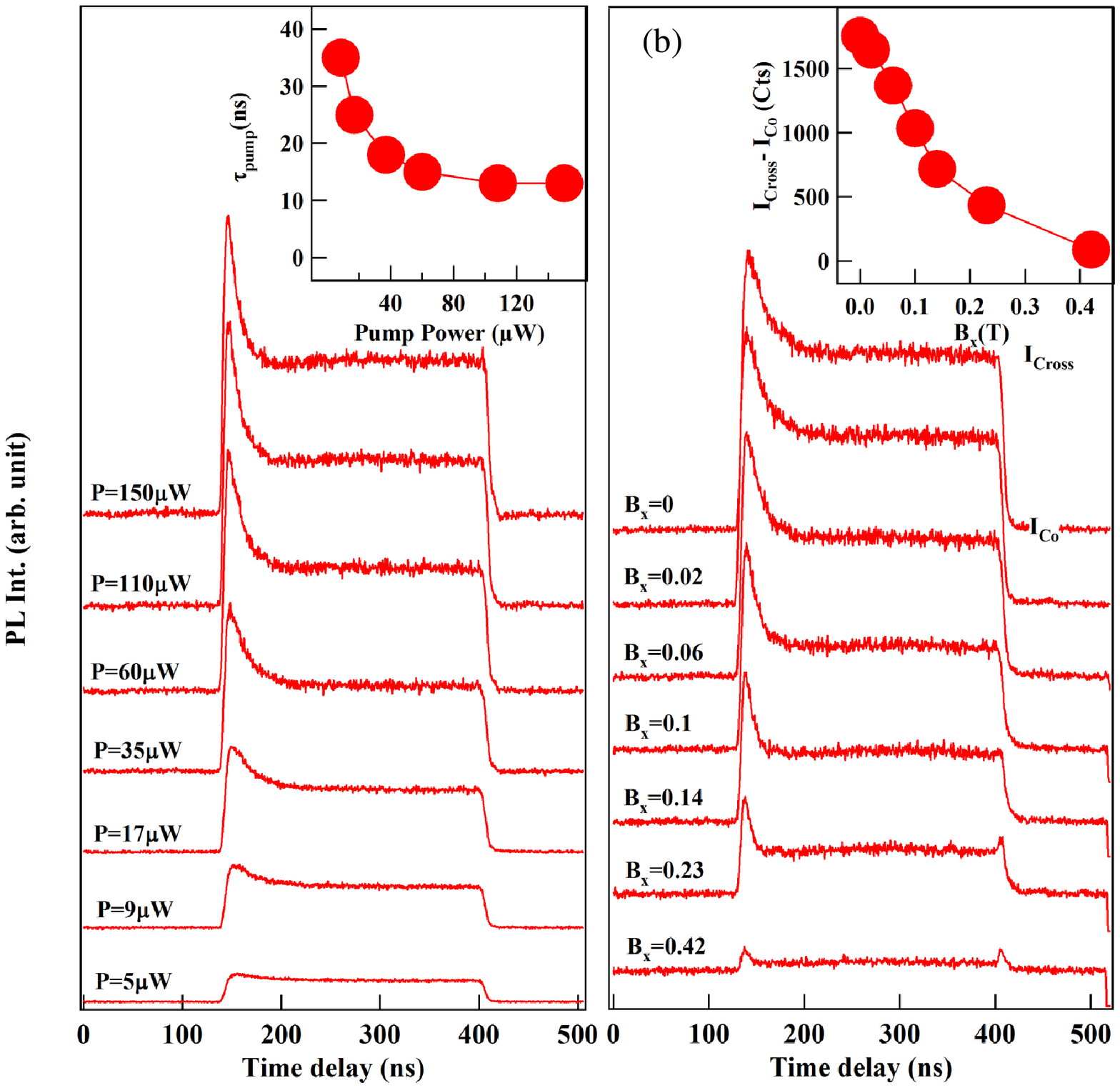}
\caption{Excitation power dependence (a) and transverse magnetic field dependence (b) of the optical pumping signal obtained for a resonant excitation on $|3,+2\rangle$. Insets: excitation power dependence of the pumping time (a) and transverse magnetic field dependence of the difference of resonant PL intensity between a $\sigma_{cross}$ and a $\sigma_{co}$ excitation (b).}
\label{Fig5}
\end{figure}

The speed of the optical pumping increases with the excitation intensity. This is presented in Fig.~\ref{Fig5}(a) in the case of a resonant excitation of $|3,\pm2\rangle$ with alternate circular polarization. At high excitation intensity, the pumping time saturates to a value similar to the width of the bunching signal observed in the auto-correlation measurements.

As observed for the auto-correlation, the resonant pumping signal is also strongly sensitive to a transverse magnetic field. Under a weak transverse field (see Fig.~\ref{Fig5}(b)), we first observe an increase of the speed of the pumping together with a decrease of the amplitude of the signal when the transient time reaches the time resolution of the set-up (around 10 ns). For a large transverse field (B$_\perp$=0.42T), the co and cross circularly polarized resonant PL intensities are identical (see the inset of Fig.~\ref{Fig5}(b)) and similar pumping transients are observed when switching from $\sigma_{co}$ to $\sigma_{cross}$ or from $\sigma_{cross}$ to $\sigma_{co}$ circular polarization.

\begin{figure}[hbt]
\includegraphics[width=3.3in]{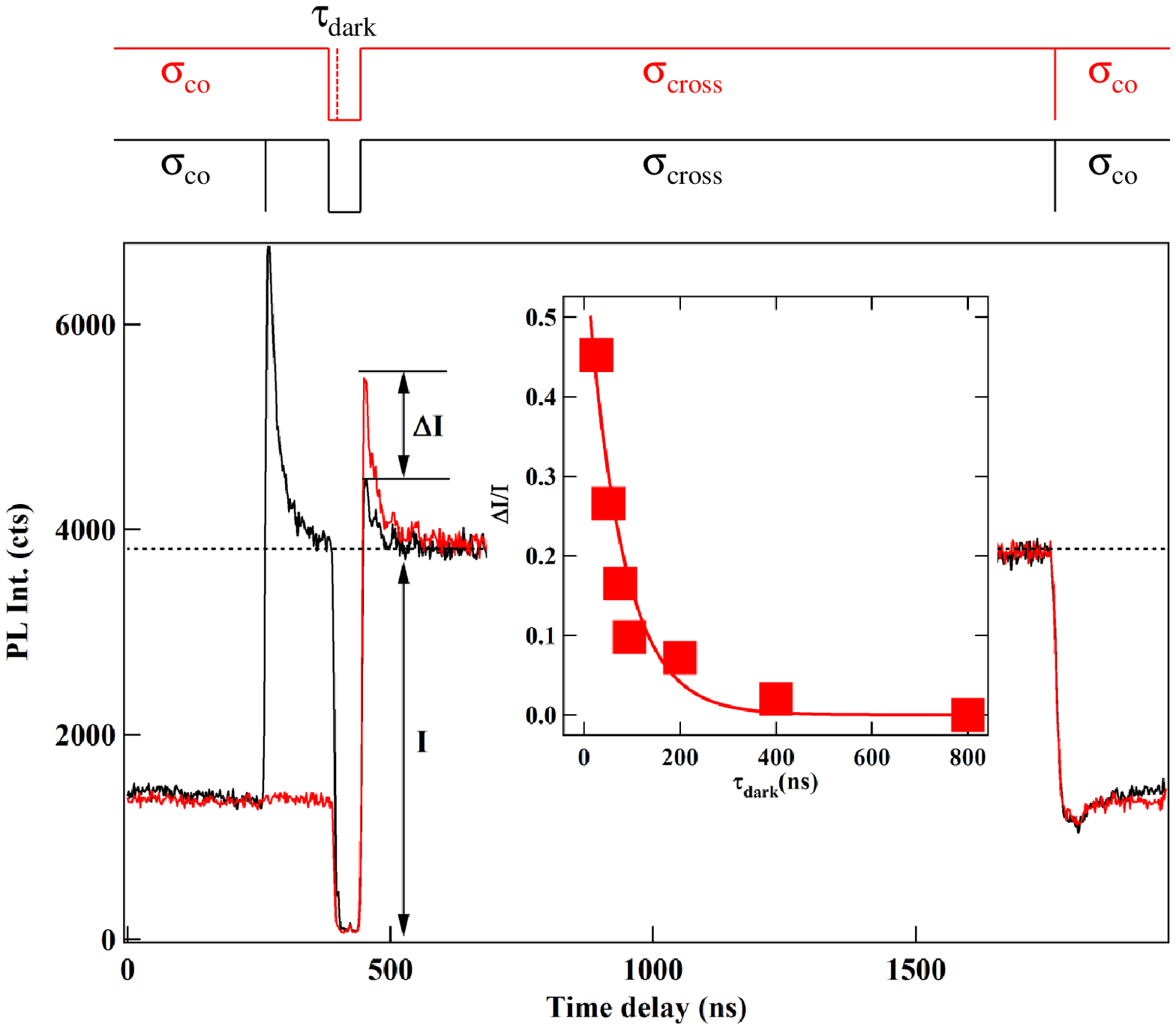}
\caption{Optical pumping for an excitation of $|3,+2\rangle$ with modulated circular polarization. A dark time ($\tau_{dark} = 50ns$) is introduced in the pumping sequence. The polarization switching of the excitation occurs either before (black) or during (red) the dark time. The black and red diagrams present the corresponding resonant excitation sequences. The inset presents the variation of the ratio $\Delta I/I$ as a function of $\tau_{dark}$. The solid line is an exponential fit with $\tau_{relax} = 80 ns$.}
\label{Fig6}
\end{figure}

To observe the relaxation of the prepared non–equilibrium distribution of the Mn-hole spins, the circularly polarized pump laser is switched off during a dark time $\tau_{dark}$. The amplitude of the pumping transient which appears after $\tau_{dark}$ depends on the Mn-hole spin relaxation. A dark time of 50 ns is enough to observe the reappearance of a significant pumping transient (Fig.~\ref{Fig6}). For comparison and for a better sensitivity of the measurement, the pumping transient observed in the absence of initial preparation of the Mn-hole spin (i.e. when switching of the circular polarization during the dark time) is also presented (red trace in Fig.~\ref{Fig6}). The normalized difference of the amplitude of these two transients, $\Delta I/I$, as a function of $\tau_{dark}$ is presented in the inset of Fig.\ref{Fig6}. This measurement shows that, when the optical excitation is off, it takes around 80 ns to the Mn-hole spin to come back to the ground state of the excited $\Lambda$ system.

If the optical pumping was storing the Mn-hole spin in the branch of the $\Lambda$ system which is not optically excited, its characteristic time would be controlled by the exciton radiative lifetime and the generation rate. With a Mn-hole relaxation time in the 100 ns range, as observed experimentally, the pumping should take place within a few nanoseconds.

Another source of spin pumping can be the leak outside the resonantly excited $\Lambda$ system. In this case, the speed of the pumping is controlled by the leakage time and, as observed experimentally, the pumping time is similar to the width of the photon bunching signal. This mechanism of pumping for the Mn-hole spin is confirmed by the transverse magnetic field dependence. The acceleration of the optical pumping in transverse magnetic field (Fig.~\ref{Fig5}(b)) has the same origin as the decrease of the width of the bunching signal. By mixing the different electron-Mn states, the transverse field enhances the leakage probability out of the resonantly driven $\Lambda$ system and decreases the corresponding optical pumping time.

\section{Modelling of the dynamics of the hybrid Mn-hole spin}

The observed large resonant PL amplitude of X$^+$-Mn and its dynamics can be qualitatively explained if a fast (nanosecond) and efficient spin transfer mechanism connects the two Mn-hole ground states of each $\Lambda$ system. Let us note that efficient Mn-hole flip-flops are also required to explain the resonant luminescence observed on neutral Mn-doped QDs (see Appendix A).

We propose a mechanism for the Mn-hole flip-flop at low temperature resulting from a deformation induced exchange interaction \cite{Tsitsishvili2003,Roszak2007}. We show here that Mn-hole states are efficiently coupled via the interplay of their exchange interaction and the lattice deformation induced heavy-hole/light-hole mixing. We will focus in the following on the two Mn-hole states $|+\frac{3}{2};\Uparrow_h\rangle$ and $|+\frac{5}{2};\Downarrow_h\rangle$ in the ground states of the $\Lambda$ system associated with the electron-Mn levels $|3,+2\rangle$ and $|2,+2\rangle$. Similar results could be obtained with the Mn-hole ground states of the other $\Lambda$ systems.

First, let us notice that the non diagonal term of the Mn-hole exchange interaction $I_{hMn}/2(S^+J^-+S^-J^+)$ couples the heavy-holes ($\Uparrow_h,\Downarrow_h)$ and light-holes $(\uparrow_h,\downarrow_h)$ levels split by $\Delta_{lh}$ through a Mn-hole flip-flop. We consider this interaction as a perturbation on the Mn heavy-hole level structure given by $I_{hMn}S_zJ_z$. To the first order in $I_{hMn}/\Delta_{lh}$, the two perturbed ground states of the $\Lambda$ system considered here $\widetilde{|+\frac{3}{2};\Uparrow_h\rangle}$ and $\widetilde{|+\frac{5}{2};\Downarrow_h\rangle}$ can be written \cite{Cohen}:

\begin{eqnarray}
\widetilde{|+\frac{5}{2};\Downarrow_h\rangle}=|+\frac{5}{2};\Downarrow_h\rangle-\frac{\sqrt{15}}{2}\frac{I_{hMn}}{\Delta_{lh}}|+\frac{3}{2};\downarrow_h\rangle\nonumber\\
\widetilde{|+\frac{3}{2};\Uparrow_h\rangle}=|+\frac{3}{2};\Uparrow_h\rangle-\frac{\sqrt{15}}{2}\frac{I_{hMn}}{\Delta_{lh}}|+\frac{5}{2};\uparrow_h\rangle
\end{eqnarray}

\noindent where we neglect the exchange energy shifts of the Mn-hole levels much smaller than $\Delta_{lh}$.

Phonon-induced deformations comes into play via the off-diagonal terms of the Bir-Pikus Hamiltonian describing the influence of strain on the valence band:

\begin{eqnarray}
\label{HBP}
H_{BP}=a_v\sum_i\epsilon_{ii}+b\sum_i\epsilon_{ii}\left(J_i^2-\frac{1}{3}J^2\right)\nonumber\\
+\frac{2d}{\sqrt{3}}\sum_{i>j}\{J_i,J_j\}\epsilon_{ij}
\end{eqnarray}

\noindent where $\epsilon_{ij}$ are the strain tensor components with $\epsilon_{ij}$=$\epsilon_{ji}$, $\{J_i,J_j\}=1/2(J_iJ_j+J_jJ_i)$ and a$_v$, b and d deformation potential constants characteristic of the material (see table \ref{paraph}). The strain produced by phonon vibrations couples the perturbed Mn-hole states $\widetilde{|+5/2\Downarrow_h\rangle}$ and $\widetilde{|+3/2\Uparrow_h\rangle}$ through the Hamiltonian term

\begin{eqnarray}
\label{int}
\widetilde{\langle+\frac{5}{2};\Downarrow_h|}H_{BP}\widetilde{|+\frac{3}{2};\Uparrow_h\rangle}=2\times(-\frac{\sqrt{15}}{2}\frac{I_{hMn}}{\Delta_{lh}})\times R^*
\end{eqnarray}

\noindent with

\begin{eqnarray}
\label{P}
R=\frac{\sqrt{3}}{2}b(\epsilon_{xx}-\epsilon_{yy})-id\epsilon_{xy}
\end{eqnarray}

\noindent a deformation dependent non-diagonal term of $H_{BP}$ \cite{Tsitsishvili2003,Roszak2007}. The coupling of the Mn-hole states is a result of an interplay between the Mn-hole exchange interaction and the deformation: neither the exchange interaction nor the deformation perturbation alone can couple these states.

According to (\ref{int}), an effective Hamiltonian describing the discussed interaction mechanism with phonons in the subspace $\{|+\frac{5}{2};\Uparrow_h\rangle,|+\frac{5}{2};\Downarrow_h\rangle, |+\frac{3}{2};\Uparrow_h\rangle,|+\frac{3}{2};\Downarrow_h\rangle\}$  is

\begin{eqnarray}
\label{Hint}
H_{int}=-\sqrt{15}\frac{I_{hMn}}{\Delta_{lh}}R^*|+\frac{5}{2};\Downarrow_h\rangle\langle+\frac{3}{2};\Uparrow_h|+H.c
\end{eqnarray}

The spin decay rates from $|+\frac{3}{2};\Uparrow_h\rangle$ to $|+\frac{5}{2};\Downarrow_h\rangle$ accompanied by the emission of an acoustic phonon is then given by Fermi's golden rule

\begin{eqnarray}
\label{fermi}
\tau^{-1}&=&\frac{2\pi}{\hbar}\sum_{k}\left|\langle+\frac{5}{2};\Downarrow_h;\psi;n_k+1|H_{int}|+\frac{3}{2};\Uparrow_h;\psi;n_k\rangle\right|^2\nonumber\\
&\times&\delta(\hbar\omega_0-\hbar\omega_{k})
\end{eqnarray}

\noindent where $\hbar\omega_0$ is the energy splitting between $|+\frac{5}{2};\Downarrow_h\rangle$ and $|+\frac{3}{2};\Uparrow_h\rangle$, $n_k$ the number of phonons in mode $k$ and $\psi$ the orbital part of the hole wave function.

To evaluate the matrix element in (\ref{fermi}) we use the strain tensor components $\epsilon_{ij}$ given by

\begin{eqnarray}
\label{eps}
{
\epsilon_{ij}=\frac{1}{2}\left(\frac{\partial u_i}{\partial r_j}+\frac{\partial u_j}{\partial r_i}\right)
}
\end{eqnarray}

\noindent where $\overrightarrow{u}(\overrightarrow{r})$ is the local displacement field. For an acoustic phonon, the quantized displacement field can be written in the real space \cite{Roszak2007,Mahan}:

\begin{eqnarray}
\label{u}
{
\overrightarrow{u}(\overrightarrow{r})=i\sum_{k,\lambda}\sqrt{\frac{\hbar}{2\rho\omega_{k,\lambda}N\nu_0}}\overrightarrow{e}_{k,\lambda}(b_{k,\lambda}+b^\dag_{-k,\lambda})e^{i\overrightarrow{k}\overrightarrow{r}}
}
\end{eqnarray}

\noindent where N is the number of unit cells in the crystal, $\nu_0$ is the volume of a cell and $\rho$ the mass density. $b^\dag_{k,\lambda}$ ($b_{k,\lambda}$) is the creation (annihilation) operator of phonon in the mode $(k,\lambda)$ of energy $\hbar\omega_{k\lambda}$ and unit polarization vector $\overrightarrow{e}_{k,\lambda}$. In zinc-blend crystals there are two transverse acoustic phonon branches $\lambda=t_1, t_2$ and one longitudinal acoustic phonon branch $\lambda=l$. The polarization vectors of these phonons branches are given by \cite{Woods2004}

\begin{eqnarray}
\label{k}
\overrightarrow{e}_{k,l}&=&\frac{\overrightarrow{k}}{k}=\frac{1}{k}(k_x,k_y,k_z)\nonumber\\
\overrightarrow{e}_{k,t_1}&=&\frac{1}{kk_{\bot}}(k_xk_z,k_yk_z,-k_{\bot}^2)\nonumber\\
\overrightarrow{e}_{k,t_2}&=&\frac{1}{k_{\bot}}(k_y,-k_x,0)
\end{eqnarray}

\noindent with $k_{\bot}=\sqrt{k_x^2+k_y^2}$.

Upon substitutions given by (\ref{eps}), (\ref{u}) and (\ref{k}), we obtain for the matrix element in (\ref{fermi}):

\begin{eqnarray}
&&|M_{k,\lambda}|^2=15\left(\frac{I_{hMn}}{\Delta_{lh}}\right)^2\frac{\hbar}{2\rho\omega_{k,\lambda}N\nu_0}\left(n_B(\omega_{k,\lambda})+1\right)\nonumber\\
&\times&\left(\frac{3b^2}{4}(k_xe_{x,\lambda}-k_ye_{y,\lambda})^2+\frac{d^2}{4}(k_xe_{y,\lambda}+k_ye_{x,\lambda})^2\right)\nonumber\\
&\times&|\mathcal{F}_{\lambda}(\overrightarrow{k})|^2
\end{eqnarray}

\noindent with

\begin{eqnarray}
\mathcal{F}_{\lambda}(\overrightarrow{k})=\int_{-\infty}^{\infty}d^3r\psi^*(\overrightarrow{r})e^{i\overrightarrow{k}\overrightarrow{r}}\psi(\overrightarrow{r})
\end{eqnarray}

\noindent and $n_B(\omega_{k,\lambda})=
1/(e^{\hbar\omega_{k,\lambda}/K_BT}-1)$, the thermal phonon distribution function.

For a Gaussian hole wave function with in-plane and z-direction parameters l$_{\bot}$ and l$_z$ respectively (full width at half maximum $2\sqrt{2\ln2}l_i$)

\begin{eqnarray}
\psi(\overrightarrow{r})=\frac{1}{\pi^{3/4}l_{\bot}\sqrt{l_z}}e^{-\frac{1}{2}\left(\left(\frac{r_{\bot}}{l_{\bot}}\right)^2+\left(\frac{z}{l_z}\right)^2\right)}
\end{eqnarray}

\noindent the form factor $\mathcal{F}_{\lambda}(\overrightarrow{k})$, which is the Fourier transform of $|\psi(\overrightarrow{r})|^2$, becomes

\begin{eqnarray}
\mathcal{F}_{\lambda}(\overrightarrow{k})=e^{-\frac{1}{4}\left(\left(l_{\bot}k_{\bot}\right)^2+\left(l_zk_z\right)^2\right)}
\end{eqnarray}

\begin{table}[hbt] \centering
\caption{Material (CdTe or ZnTe) \cite{Adachi2005} and QD parameters used in the calculation of the coupled hole and Mn spin relaxation time.}
\begin{tabular}{lcr}
\hline\hline
CdTe& &\\
\hline
Deformation potential constants & $|b|$ &  1.0 eV  \\
& $|d|$ &  4.4 eV  \\
Longitudinal sound speed & c$_l$ &  3300 m/s  \\
Transverse sound speed & c$_t$ &  1800 m/s  \\
Density & $\rho$ &  5860 kg/m$^3$  \\
\hline
ZnTe& &\\
\hline
Deformation potential constants & $|b|$ &  1.4 eV  \\
& $|d|$ &  4.4 eV  \\
Longitudinal sound speed & c$_l$ &  3800 m/s  \\
Transverse sound speed & c$_t$ &  2300 m/s  \\
Density & $\rho$ &  5908 kg/m$^3$  \\
\hline
Quantum dot& &\\
\hline
Mn-hole exchange energy & I$_{hMn}$ &  0.35 meV  \\
hh-lh exciton splitting&  $\Delta_{lh}$&  15 meV  \\
Hole wave function widths: & &   \\
- in plane & l$_{\bot}$ &3.0 nm   \\
- z direction  & l$_z$ &1.25 nm   \\
\hline\hline
\end{tabular}
\label{paraph}
\end{table}

Considering a linear dispersion of acoustic phonons $\omega_{k,\lambda}=c_{\lambda}k$ and in spherical coordinates with $\overrightarrow{k}=k(\sin\theta\cos\varphi,\sin\theta\sin\varphi,\cos\theta)$, the explicit formula of the decay rate (\ref{fermi}) is

\begin{eqnarray}
\tau^{-1}&=&\sum_{\lambda}\frac{15}{(2\pi)^2}\left(\frac{I_{hMn}}{\Delta_{lh}}\right)^2\left(\frac{\omega_0}{c_{\lambda}}\right)^3\frac{1}{2\hbar\rho c_{\lambda}^2}\frac{\pi}{4}\left(3b^2+d^2\right)\nonumber\\
&\times&\left(n_B(\omega_0)+1)\right)\int_0^{\pi}d\theta\sin\theta|\mathcal{F}_{\lambda}(\omega_0,\theta)|^2G_{\lambda}(\theta)
\end{eqnarray}

\noindent where we used the continuum limit ($\sum_k\rightarrow V/(2\pi)^3\int d^3k$ with $V=N\nu_0$ the crystal volume) and integrated over $k$ and $\varphi$. The summation is taken over the acoustic phonon branches $\lambda$ of corresponding sound velocity c$_{\lambda}$. The geometrical form factors for each phonon branch, $G_{\lambda}(\theta)$, are given by

\begin{eqnarray}
G_{l}(\theta)&=&\sin^4\theta\nonumber\\
G_{t_1}(\theta)&=&\sin^2\theta\cos^2\theta\nonumber\\
G_{t_2}(\theta)&=&\sin^2\theta
\end{eqnarray}

\begin{figure}[hbt]
\includegraphics[width=3.3in]{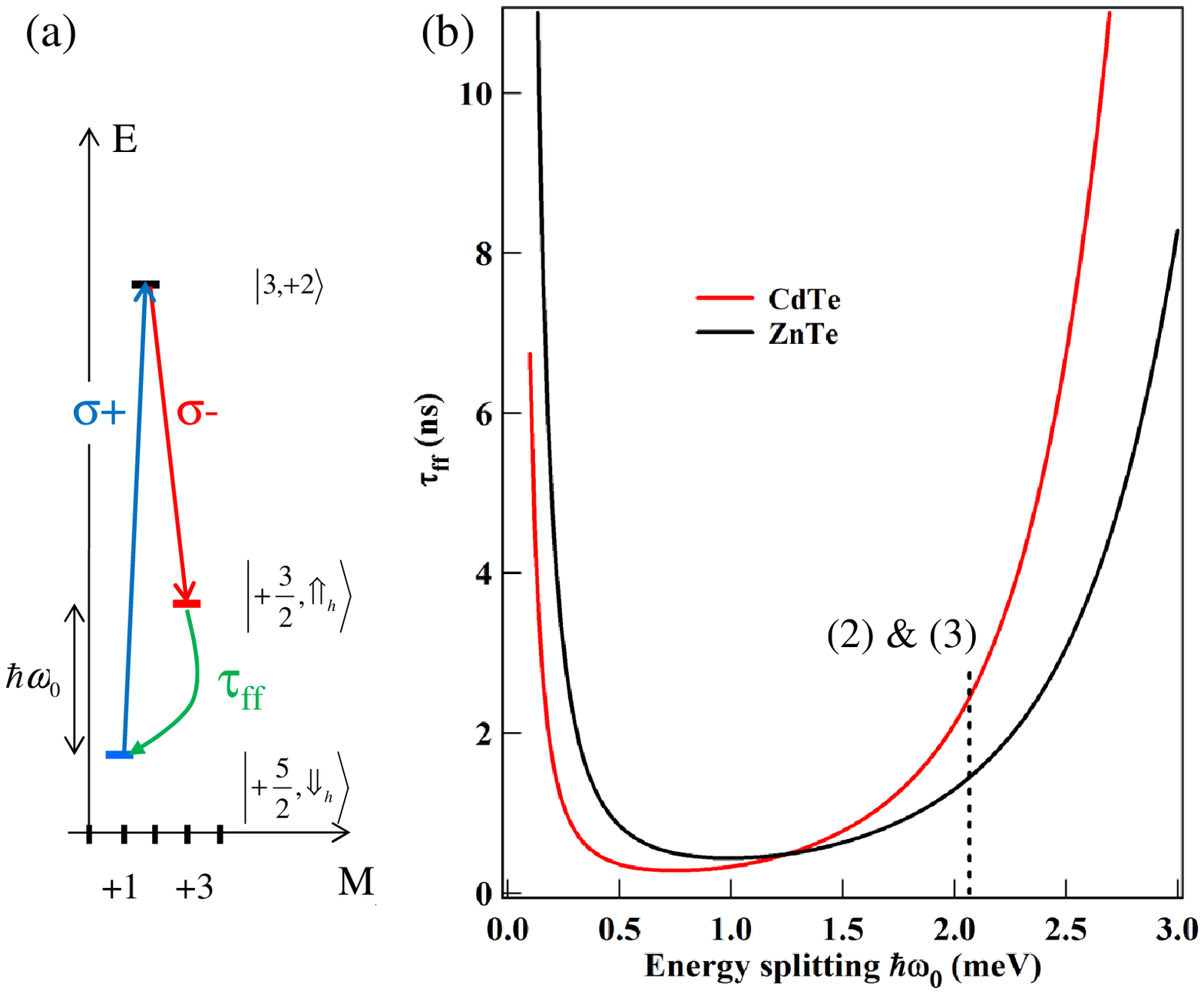}
\caption{(a) Scheme of the energy levels of the optical $\Lambda$ system associated with the electron-Mn state $|3,+2\rangle $ extracted from the full level structure of a positively charged Mn-doped QD (Fig. \ref{Fig1}(b)). (b) Relaxation time $\tau_{ff}$, between the two Mn-hole ground states of the $\Lambda$ system  calculated with the material and QD parameters listed in table \ref{paraph} and a temperature T=7K. The vertical line shows the energy splitting in the studied QD of the Mn-hole states involved in the $\Lambda$ systems considered here (Resonances (2) and (3) identified in Fig.~\ref{Fig2}).}
\label{Fig7}
\end{figure}

In the numerical calculation of the spin flip time $\tau_{ff}$ presented in Fig.\ref{Fig7} we use the material parameters of CdTe or ZnTe and the typical parameters for self-assembled CdTe/ZnTe QDs listed in Table \ref{paraph}. The calculated relaxation time strongly depend on the energy separation between the Mn-hole levels $\hbar\omega_0$. This energy dependence is controlled by the size of the hole wave-function given by l$_{\bot}$ and l$_z$. The estimated flip-flop time is also strongly sensitive on the exchange induced mixing of the ground heavy-hole states with the higher energy light-hole levels. In our model, this mixing is controlled by $\Delta_{lh}$, an effective energy splitting between heavy-holes and light-holes. This simple parameter can indeed describe more complex effects such as a coupling of the confined heavy-hole with ground state light-holes in the barriers \cite{Michler2003} or effective reduction of heavy-hole/light-hole splitting due to a presence of a dense manifold of heavy-hole like QD states lying between the confined heavy-hole and light-hole levels \cite{Bester2015}. From this modelling we deduce that for a hole confined in small Cd$_x$Zn$_{1-x}$Te alloy QDs, the Mn-hole flip-flop time $\tau_{ff}$ can be easily bellow 2 ns for an effective heavy-hole/light-hole splitting $\Delta_{lh}$=15 meV and an energy separation in the meV range, typical for the Mn-hole spin splitting in magnetic QDs. We will use in the following calculations $\tau_{ff}$=1.5 ns for the ground states of each $\Lambda$ system.

\section{Model of the carrier-Mn spin dynamics under resonant excitation}

Using the level scheme presented in Fig.\ref{Fig1}(b) for a positively charged Mn-doped QD and the estimated Mn-hole flip-flop rates, we can calculate the time evolution of the 24x24 density matrix $\varrho$ describing the population and the coherence of the 12 electron-Mn states on the excited state and the 12 Mn-hole states on the ground state of a positively charged QD. In the Markovian approximation, the master equation which governs the evolution of $\varrho$ can be written in a general form (Lindblad form) as:

\begin{equation}
\label{Lindblad} {\frac{\partial\varrho}{\partial t}=\frac{-i}{\hbar}[{\cal H},\varrho]+L\varrho}
\end{equation}

\noindent where ${\cal H}$ is the Hamiltonian of the complete system ($X^+$-Mn ${\cal H}_{X^+Mn}$ and Mn-hole ${\cal H}_{hMn}$):

\begin{eqnarray}
{\cal H}_{X^+Mn}=I_{eMn}\vec{S}\cdot\vec{\sigma}-2\eta S_z^2+D_0S^2_z+E(S_y^2-S_x^2)\nonumber\\
+g_{Mn}\mu_B\vec{S}\cdot\vec{B}+g_{e}\mu_B\vec{\sigma}\cdot\vec{B}
\end{eqnarray}

\noindent and

\begin{eqnarray}
{\cal H}_{hMn}=I_{hMn}\vec{S}\cdot\vec{J}-\eta S_z^2+D_0S^2_z+E(S_y^2-S_x^2)\nonumber\\
+g_{Mn}\mu_B\vec{S}\cdot\vec{B}+g_{h}\mu_B\vec{J}\cdot\vec{B}
\end{eqnarray}

In (\ref{Lindblad}), $L\varrho$ describes the coupling or decay channels resulting from an
interaction with the environment \cite{Exter2009, Roy2011, Jamet2013}. The population transfers from level $j$ to level $i$ in an irreversible process associated with a coupling to a reservoir is described by a Lindblad term of the form

\begin{eqnarray}
\label{inc}
L_{inc,j\rightarrow i}\varrho=\frac{\Gamma_{j\rightarrow i}}{2}(2|i\rangle\langle j|\varrho|j\rangle\langle i| -\varrho|j\rangle\langle j|-|j\rangle\langle j|\varrho)
\end{eqnarray}

\noindent where $\Gamma_{j\rightarrow i}$ is the incoherent relaxation rate from level $j$ to level $i$. Such term can describe the radiative decay of the exciton (irreversible coupling to the photon modes) or the relaxation of the carriers or Mn spin (irreversible coupling to the phonon modes). It can also be used to describe the optical generation of an exciton in the low excitation regime where the energy shift induced by the coupling with the laser field is neglected.

A pure dephasing (i.e. not related to an exchange of energy with a reservoir) can also be introduced for the different spins and described by $L_{deph,jj}\varrho$:

\begin{eqnarray}
\label{deph} {L_{deph,jj}\varrho=\frac{\gamma_{jj}}{2}(2|j\rangle\langle j|\varrho|j\rangle\langle j| -\varrho|j\rangle\langle j|-|j\rangle\langle j|\varrho)}
\end{eqnarray}

\noindent where $\gamma_{jj}$ is the pure dephasing rate of level $j$.

To identify the main spin relaxation channels responsible for the observed spin fluctuations, we first modelled the auto-correlation of the resonant PL using the full spin level structure of a p-doped magnetic QD. For a qualitative description of the observed spin dynamics, we use as an example the Mn-doped QD parameters extracted from the linear polarization intensity map listed in table \ref{paraQD} and reasonable order of magnitude for the spin relaxation times.

As already observed in charged Mn-doped QDs under pulsed resonant excitation (ref. 23), we consider that the spin dynamics in the excited state is controlled by the time evolution of ${\cal H}_{X^+Mn}$, the generation rate of excitons $\gamma_g$=1/$\tau_g$ and their radiative lifetime $\tau_r$=0.3 ns. The coherence of the coupled electron-Mn spins is limited by a pure dephasing term, $T_2^{eMn}$=0.5 ns, typical value measured in charged Mn-doped QDs\cite{Lafuente2015}.

For the Mn-hole system in the ground state, we take into account a spin relaxation time of the Mn in the exchange field of the hole, $\tau_{Mn}$, describing relaxation channels involving a change of the Mn spin by one unit. This spin relaxation channel is introduced for a general description, however its characteristic time (in the $\mu$s range) is long compared to the time-scale of the dynamics considered in the resonant PL experiments and does not qualitatively affect the calculated time evolution.

Because of the presence of valence band mixing in the QDs, spin flip of the hole independently of the Mn are expected to be more efficient. A spin flip time in the 10 ns range has indeed been calculated for a hole in the exchange field of a Mn \cite{Cao2011,Cywinski2010}. Relaxation time of the hole spin around 5 ns has also been measured at zero magnetic field in negatively charged CdTe/ZnTe QDs \cite{LeGall2012}. We then include in the model possible spin flips of the hole by one unit with a characteristic time $\tau_{h}$=10ns. The phonon induced Mn-hole flip-flops, occurring at $\tau_{ff}$, are also introduced between the two Mn-hole ground states of each $\Lambda$ system.

For a general qualitative description, an additional pure dephasing time $T_2^{hMn}$ is also included in the dynamics of the Mn-hole system with a Lindblad term of the form (\ref{deph}). We cannot extract this parameter from the experiments. We take $T_2^{hMn}$= 5 ns, slightly longer than what was measured for electron-Mn, as the Mn-hole system is highly split and less sensitive to effective fluctuating magnetic field such as the one produced by nuclear spins for instance \cite{LeGall2012,Houel2014}.

\begin{figure}[hbt]
\includegraphics[width=3.3in]{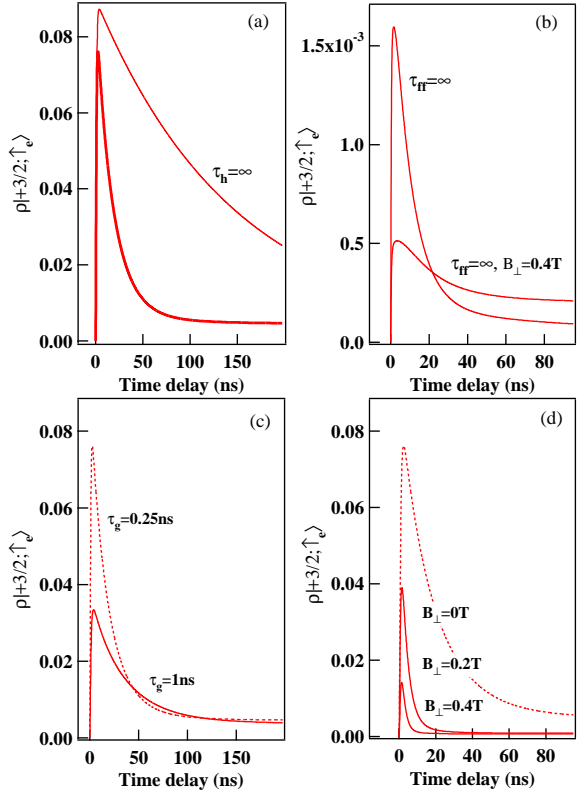}
\caption{(a) Calculated time evolution of $\rho_{|+\frac{3}{2},\uparrow_e\rangle}$ with the QD parameters listed in table \ref{paraQD} and (unless specified) $\tau_r$=0.3 ns, $\tau_{Mn}$=5 $\mu$s, $\tau_h$=10 ns, $\tau_g$=0.25 ns, $\tau_{ff}$=1.5 ns, $T_2^{hMn}$= 5 ns, $T_2^{eMn}$= 0.5 ns, T=7K and B$_{\perp}$=0. (b) (c) and (d) illustrate the influence on $\rho_{|+\frac{3}{2},\uparrow_e\rangle}(t)$ of $\tau_{ff}$, $\tau_g$ and $B_{\perp}$ respectively. Note the different vertical scale in (b).}
\label{Fig8}
\end{figure}

The transition rates $\Gamma_{\gamma\rightarrow\gamma'}$ between the different Mn-hole states depend on their energy separation $E_{\gamma\gamma'}=E_{\gamma'}-E_{\gamma}$. Here we use $\Gamma_{\gamma\rightarrow\gamma'}$=1/$\tau_{i}$ if $E_{\gamma\gamma'}<0$ and $\Gamma_{\gamma\rightarrow\gamma'}$=1/$\tau_{i}e^{-E_{\gamma\gamma'}/k_BT}$ if $E_{\gamma\gamma'}>0$ \cite{Govorov2005,Cao2011}. This accounts for a thermalization among the 12 Mn-hole levels with an effective spin temperature $T$. The optical excitation ($\tau_g$), the exciton recombination ($\tau_r$), the Mn spin relaxation ($\tau_{Mn}$), the hole spin relaxation ($\tau_{h}$) and the phonon induced transfer time ($\tau_{ff}$) produce a irreversible population transfer between level $\gamma$ and $\gamma'$ and are described by Lindblad terms (\ref{inc}).

\begin{figure}[hbt]
\includegraphics[width=3.3in]{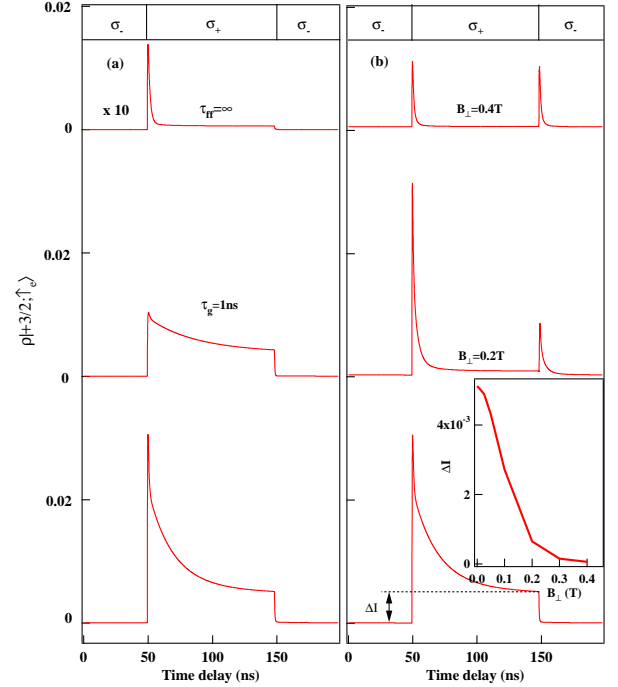}
\caption{Calculated resonant optical pumping transients for a $\sigma-$ detection and an excitation of $|3,+2\rangle$ and $|3,-2\rangle$ with modulated circular polarization. The QD parameters for the calculations are those listed in table \ref{paraQD} and $\tau_r$=0.3 ns, $\tau_{Mn}$=5 $\mu$s, $\tau_h$=10 ns, $T_2^{hMn}$= 5 ns, $T_2^{eMn}$= 0.5 ns, $\tau_{ff}$=1.5 ns, T=10 K and $\tau_g$=0.25 ns. (a) Influence of a variation of $\tau_g$ and $\tau_{ff}$. (b) Influence of a transverse magnetic field $B_{\perp}$. The inset presents the transverse magnetic field dependence of the difference of population for a $\sigma+$ or a $\sigma-$ excitation.}
\label{Fig9}
\end{figure}

To model the auto-correlation of the $\sigma-$ PL intensity of the electron-Mn state $|3;+2\rangle$ under $cw$ $\sigma+$ resonant excitation we calculate the time evolution of $\rho_{|+\frac{3}{2};\uparrow_e\rangle}(t)$ with the initial condition $\rho_{|+\frac{3}{2};\Uparrow_h\rangle}(0)=1$ corresponding to the Mn-hole spin in the state $|+\frac{3}{2};\Uparrow_h\rangle$ just after the emission of a $\sigma-$ photon on the low energy branch of the $\Lambda$ system. This initial state is a slight approximation: in the presence of valence band mixing, the two ground states of a given $\Lambda$ system are not completely pure Mn-hole spin states but are slightly coupled by a Mn-hole flip-flop induced by the exchange interaction ${\cal H}_{hMn}^{ex}$. However, as the splitting between the states $|+\frac{3}{2};\Uparrow_h\rangle$ and $|+\frac{5}{2};\Downarrow_h\rangle$ ($\Delta=4\times3/2I_{hMn}$) is large compared with the coupling term ($W=\sqrt{15}\frac{\rho_c}{\Delta_{lh}}I_{hMn}$), their coherent coupling is weak. With a large valence band mixing $\frac{\rho_c}{\Delta_{lh}}=0.1$ as observed in the dot discussed in this paper, this leads for the Mn-hole system initialized in the state $|+\frac{3}{2};\Uparrow_h\rangle$ to a fast oscillation of the population between the two corresponding Mn-hole ground states of the $\Lambda$ system with a maximum amplitude of about 1.6\% and an average population transfer efficiency of 0.8\% \cite{Cohen}. Under resonant excitation on the high energy branch of the the $\Lambda$ system, the QD remains OFF more than 99$\%$ of the time. As we will see in the following, the contribution of this weak coherent population transfer to the calculated auto-correlation signal is not significant.

$\rho_{|+\frac{3}{2};\uparrow_e\rangle}(t)$ obtained with the QD parameters listed in table \ref{paraQD} is presented in Fig.~\ref{Fig8}(a). This quantity has to be normalized by $\rho_{|+\frac{3}{2};\uparrow_e\rangle}(\infty)$) to directly account for the autocorrelation signal. After a fast increase, the calculated population presents a maximum at short delay. This model is based on a large number of parameters, whose values cannot all be extracted precisely from the measurements however, with reasonable spin relaxation parameters (see details in the caption of Fig.~\ref{Fig8}), the width and the amplitude of the maximum are in good agreement with the photon bunching signals observed experimentally.

The width of the calculated bunching is controlled by all the spin-flip terms that can induce an escape out of the resonantly excited $\Lambda$ system. At zero transverse magnetic field, it is dominated by spin flips in the Mn-hole system. As illustrated in Fig.~\ref{Fig8}(a), suppressing $\tau_h$ gives a width of bunching only controlled by the Hamiltonian evolution and the decoherence which is slightly larger than what is observed experimentally (Fig.~\ref{Fig3}).

The dependence on the excitation intensity, $\tau_g$, and transverse magnetic field, $B_{\perp}$, are also qualitatively well reproduced by the model (Fig.~\ref{Fig8}(c) and (d) respectively). At zero magnetic field, the leaks outside the excited $\Lambda$ systems are dominated by $\tau_h$. $\mathcal{H}_{X^+Mn}$ induces fluctuations in a slightly longer time scale. The situation is different under a weak transverse magnetic field where the electron-Mn states are mixed introducing new channel of escape and significantly reducing the width of the photon bunching (See Fig.~\ref{Fig3} for the corresponding experiments).

Let us note that suppressing the fast flip-flop process connecting the two Mn-hole ground states ($\tau_{ff}=\infty$ in Fig.~\ref{Fig8}(b)) still produces a bunching as with the approximated initial condition used in the calculation ($\rho_{|+\frac{3}{2};\Uparrow_h\rangle}(0)=1$) a weak coherent transfer between the two ground states of the $\Lambda$ system still exist. However, with this process only, the calculated PL intensity is always more than 50 times smaller than with $\tau_{ff}$ and its contribution to the calculated auto-correlation signal (Fig.~\ref{Fig8}(a)) can be safely neglected.

With this model, we can also calculate the population of the electron-Mn states under resonant excitation with alternated circular polarization and estimate the efficiency and dynamics of the optical pumping. Figure~\ref{Fig9} presents the calculated time evolution of the population of the electron-Mn state $|+\frac{3}{2},\uparrow_e\rangle$ under alternated resonant excitation of $|3,+2\rangle$ in $\sigma+$ polarization or $|3,-2\rangle$ in $\sigma-$ polarization. This corresponds to the experimental configuration where the QD is resonantly excited with modulated circular polarization at the energy of $|3,+2\rangle$ and $|3,-2\rangle$ (absorption (2) in Fig.~\ref{Fig2}(b)) and the low energy resonant PL is detected in $\sigma-$ polarisation. The main features of the time-resolved optical pumping experiments (see Fig.~\ref{Fig4} and Fig.~\ref{Fig5}) are well reproduced by the model. The timescale of the pumping transient, in the few tens of nanosecond range, and its excitation intensity dependence are also in good agrement with the experiments (see figure~\ref{Fig9}(a)).

The influence of a transverse magnetic field, B$_{\perp}$, on the optical pumping transient can also be described by this model. First, a significant reduction of the pumping time is observed for a weak magnetic field (B$_{\perp}$=0.2T in Fig.~\ref{Fig9}(b)). As for the autocorrelation, this acceleration comes from the increase of the leakage out of the $\Lambda$ system induced by the mixing of the electron-Mn states. Secondly, the transients obtained when switching the polarization from $\sigma_{co}$ to $\sigma_{cross}$ and from $\sigma_{cross}$ to $\sigma_{co}$ become identical for B$_{\perp}\approx0.4$T, as observed in the experiments (Fig.~\ref{Fig5}(b)).

To understand this behaviour under B$_{\perp}$, let us remember that we resonantly excite $|3,+2\rangle$ from $|+\frac{5}{2},\Downarrow_h\rangle$ with $\sigma+$ light and excite $|3,-2\rangle$ from $|-\frac{5}{2},\Uparrow_h\rangle$ with $\sigma-$ photons. In both cases we detect the population of $|+\frac{3}{2},\uparrow_e\rangle$ in $\sigma-$ polarization (see the excitation/detection configuration illustrated in the inset of Fig.~\ref{Fig4}(b)). If the states $|3,+2\rangle$ and $|3,-2\rangle$ are uncoupled, as it is the case at zero field, we do not detect any light during the $\sigma-$ excitation. With a sufficiently large mixing of $|3,+2\rangle$ and $|3,-2\rangle$ induced by the transverse magnetic field, for a $\sigma-$ excitation of $|3,-2\rangle$, the population can be coherently transferred to $|3,+2\rangle$ during the charged exciton lifetime and $\sigma-$ light is detected after a recombination towards $|+\frac{3}{2},\Uparrow_h\rangle$ \cite{Lafuente2015}. In the optical pumping sequence, we can then observe, in $\sigma-$ polarization, a transient when the $\sigma+$ excitation empties the state $|+\frac{5}{2},\Downarrow_h\rangle$ but also a similar transient when the $\sigma-$ excitation empties the state $|-\frac{5}{2},\Uparrow_h\rangle$. The transverse magnetic field dependence of the difference of steady state intensity observed in $\sigma_{co}$ and $\sigma_{cross}$ polarization (inset of Fig.~\ref{Fig5}(b)) is also well reproduced by the model (inset of Fig.~\ref{Fig9}(b)). This depolarization curve is controlled by the anisotropy of the electron-Mn spin which is induced by $\eta$ and $D_0$ \cite{Varghese2014}. Let us note finally that, as expected, suppressing $\tau_{ff}$ from the model, a very weak average resonant PL and a fast optical pumping are obtained (Fig.~\ref{Fig9}(a), top curve).

\begin{figure}[hbt]
\includegraphics[width=3.3in]{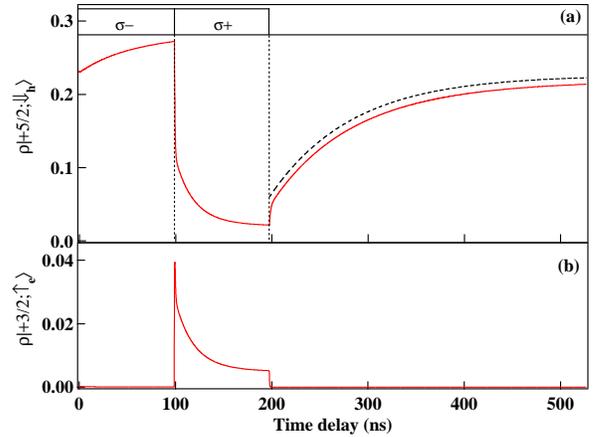}
\caption{(a) Calculated time evolution in the dark of the population of the Mn-hole state $|+\frac{5}{2},\Downarrow_h\rangle$ initialized by a sequence of $\sigma-$/$\sigma+$ resonant excitation of $|3,-2\rangle$ and $|3,+2\rangle$. The dashed black line (shifted for clarity) is an exponential fit with a characteristic time $\tau_{relax}$=85 ns. (b) Corresponding calculated time evolution of the population of $|+\frac{3}{2},\uparrow_e\rangle$. The parameters are those of Fig.~\ref{Fig9}.}
\label{Fig10}
\end{figure}

Including a dark time in the pumping sequence, we can also numerically evaluate the time required for the Mn-hole spin to return to the ground state of the excited $\Lambda$ system. The time evolution of the population of the Mn-hole state $|+5/2,\Downarrow_h\rangle$ initially prepared by a sequence of $\sigma-$/$\sigma+$ excitation resonant with $|3;+2\rangle$ (and $|3;-2\rangle$) is presented in Fig.\ref{Fig10}. When the optical excitation is switched off, after an abrupt jump due to the optical recombination of the charge exciton, the ground Mn-hole state $|+5/2,\Downarrow_h\rangle$ is repopulated in a timescale of about 100 ns, much shorter than the Mn spin relaxation time used in the model ($\tau_{Mn}$=5ns). This relaxation is induced by the presence of valence band mixing. In the presence of valence-band mixing, $\mathcal{H}_{hMn}^{ex}$ couples two by two the different Mn-hole levels. This coupling induces a transfer of population between the different Mn-hole levels. The transfer of population becomes irreversible in the presence of dephasing and controls the observed Mn-hole spin relaxation \cite{Varghese2014}.

\begin{figure}[hbt]
\includegraphics[width=3.3in]{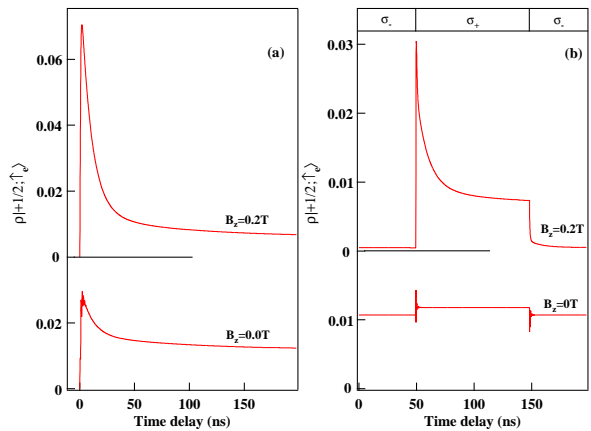}
\caption{(a) Calculated time evolution of $\rho_{|+\frac{1}{2},\uparrow_e\rangle}$ with $\rho_{|+\frac{1}{2},\Uparrow_h\rangle}$=1 (Mn-hole spin in the state $|+\frac{1}{2},\Uparrow_h\rangle$ after a $\sigma-$ recombination) for a resonant $\sigma+$ excitation of the coupled electron-Mn states $|3,+1\rangle$ and $|3,-1\rangle$ without and with a longitudinal magnetic field. (b) Time evolution of $\rho_{|+\frac{1}{2},\uparrow_e\rangle}$ under excitation with modulated circular polarization. The parameters used in the calculations are those of Fig.~\ref{Fig9}.}
\label{Fig11}
\end{figure}

The particular behaviour observed for a resonant excitation of the electron-Mn states $|3,+1\rangle$ or $|3,-1\rangle$ (weak photon bunching and no optical pumping at zero field, Fig.~\ref{Fig2}(a) and Fig.~\ref{Fig4}(a) respectively) is also qualitatively explained by the model (see figure~\ref{Fig11}). As demonstrated in Ref. 25, the presence of a strain anisotropy term $E$ in the $\mu$eV range in the Mn fine structure directly couples the states $|3,+1\rangle$ and $|3,-1\rangle$ which are initially degenerated. The splitting between the two new eigenstates in the $\mu eV$ range is much weaker than width of the resonant laser used in our experiments (around 10 $\mu eV$) and the width of the optical transitions (around 50 $\mu eV$). Under circularly polarized resonant excitation we either excite $|3,+1\rangle$ with $\sigma+$ photons or $|3,-1\rangle$ with $\sigma-$ photons. At zero magnetic field, the population is transferred between the two states in a time scale of a few hundreds picoseconds \cite{Lafuente2015}. Under circularly polarized resonant excitation, the two $\Lambda$ systems associated with $|3,\pm1\rangle$ are simultaneously excited. For alternated circular polarization, a steady state is reached and no pumping transient induced by a leak outside the $\Lambda$ systems is expected. Under a weak longitudinal magnetic field the Mn Zeeman energy dominates the strain anisotropy term and the coherent transfer is blocked. The states $|3,+1\rangle$ and $|3,-1\rangle$ are decoupled and a large amplitude of bunching and an efficient optical pumping are restored. This behaviour observed in the experiments is qualitatively reproduced by the model.

Let us finally note that in the modelling of optical pumping at zero magnetic field presented in Fig.~\ref{Fig11}(b), fast oscillations are obtained in the first nanoseconds after the polarization switching. These are due to the population transfer between $|3,+1\rangle$ and $|3,-1\rangle$ in the excited state (directly coupled by E) during the coherence time. These oscillations are too fast to be observed in the experiments. The calculated resonant PL intensity in $\sigma-$ polarization (proportional to $\rho_{|+\frac{1}{2},\uparrow_e\rangle}$) is also slightly larger for a $\sigma+$ excitation than for a $\sigma-$ excitation. The $\sigma-$ resonant PL probes the population of $|3,+1\rangle$ which is directly excited by a resonant $\sigma+$ laser (see the excitation/detection configuration in the inset of Fig.~\ref{Fig4}(a)). On the other hand, under a $\sigma-$ laser, one excites $|3,-1\rangle$ and the charged exciton has a probability to recombine before being transferred to $|3,+1\rangle$ and detected in $\sigma-$ PL. This transfer time results in a slight difference in the steady state resonant PL intensity obtained in a $\sigma_{Co}$ or $\sigma_{Cross}$ configuration (see Fig.~\ref{Fig4}(a)).

The model we developed here explains the main behaviour of a positively charged Mn-doped QD under resonant optical excitation. It shows in particular the importance of the effective heavy-hole/light-hole splitting $\Delta_{lh}$ on the Mn-hole spin dynamics. The studied CdTe/ZnTe QDs have a weak valence band offset. The resulting small value of $\Delta_{lh}$ is first responsible for the large influence of the QDs' shape or strain anisotropy on the valence band mixing. The valence band mixing reduces the magnetic anisotropy of the Mn-hole system and its spin life time. A small $\Delta_{lh}$ also significantly enhances the coupling of the Mn-hole spin with acoustic phonons and the phonon induced Mn-hole flip-flops $\tau_{ff}$. This fast spin dynamics limits the use of such hybrid spin system in practical quantum information devices. The use of different QD systems with a larger valence band offset and a larger heavy-hole/light-hole splitting \cite{Moehl2004} should significantly slow down the Mn-hole spin dynamics.

\section{Conclusion}

Using resonant PL of the positively charged exciton, we have identified an efficient spin relaxation channel for the hybrid Mn-hole spin in a QD. A modelling confirms that Mn-hole flip-flops in a nanosecond timescale are induced by an interplay of Mn-hole exchange interaction and the lattice deformation of acoustic phonons. These flip-flops are responsible for the large PL intensity observed under resonant excitation of the $\Lambda$ systems present in positively charged Mn-doped QDs. We showed that jumps out of an optically excited $\Lambda$ system are possible. These leaks induce a large bunching of the resonant PL and are at the origin of the optical pumping of the Mn-hole spin observed under circularly polarized resonant excitation. Escape out of the excited $\Lambda$ system can be enhanced by a transverse magnetic field which mixes electron-Mn states in the excited state of the charged QD. The fast Mn-hole spin dynamics revealed by these experiments make difficult the practical use of this hybrid spin for quantum information devices. However, the use of a different QD systems with a better hole confinement and a larger heavy-hole/light-hole splitting would significantly reduce the influence of valence band mixing on the Mn-hole spin relaxation and limit the interaction of the hybrid Mn-hole spin with acoustic phonons.

\begin{acknowledgements}

This work was realized in the framework of the Commissariat \`{a} l'Energie Atomique et aux Energies Alternatives (Institut Nanosciences et Cryog\'{e}nie) / Centre National de la Recherche Scientifique (Institut N\'{e}el) joint research team NanoPhysique et Semi-Conducteurs.

\end{acknowledgements}

\begin{appendix}

\section{Mn-hole flip-flops observed in neutral magnetic quantum dots.}

\begin{figure}[hbt]
\includegraphics[width=3.3in]{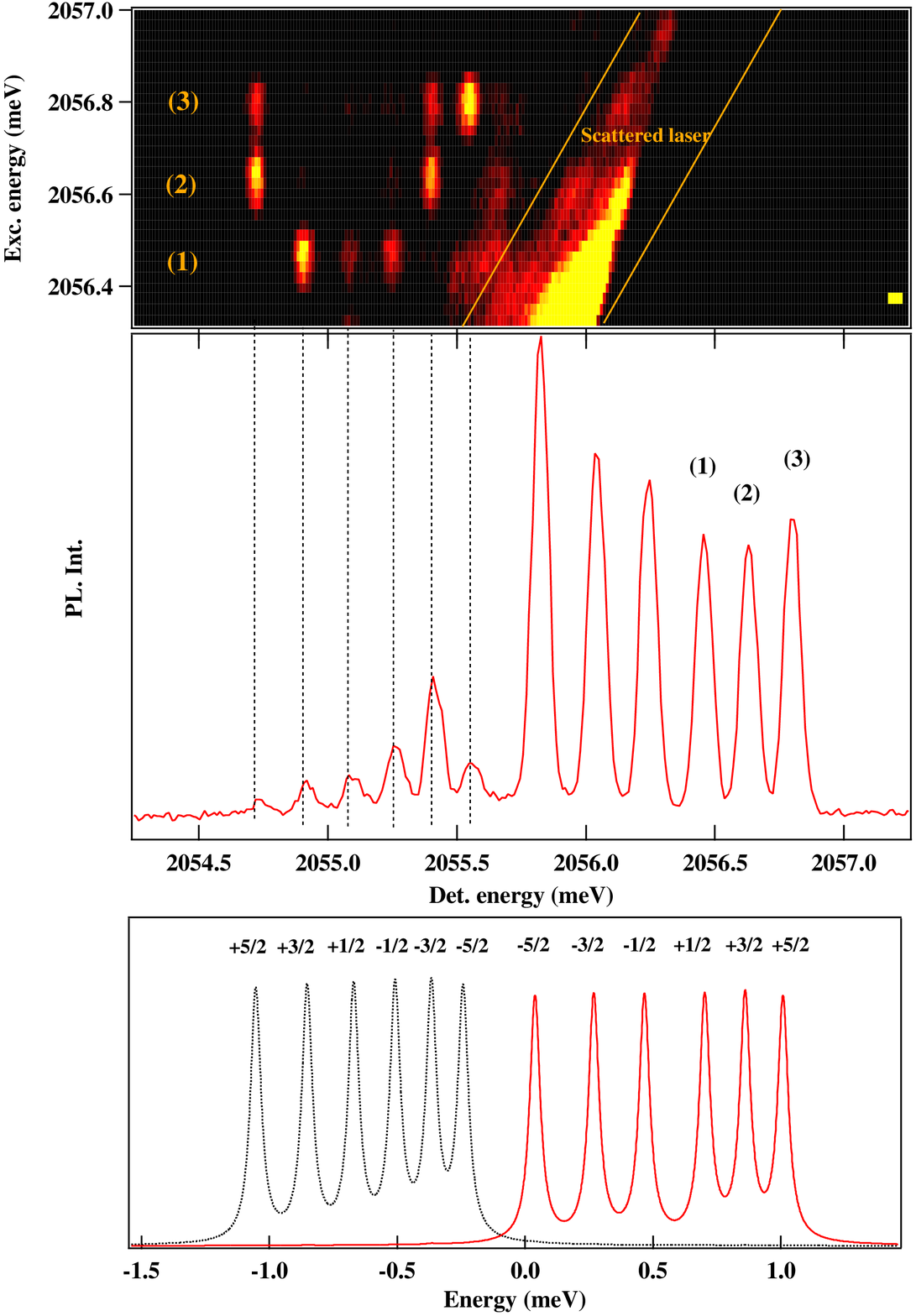}
\caption{PL intensity map detected on the dark exciton states of a Mn-doped QD. The excitation laser is scanned across the three high energy bright exciton states. Excitation and detection are circularly cross-polarized. The resonances (1-3) are discussed in the text. The bottom panel presents bright $|+1\rangle$ (plain line) and dark $|-2\rangle$ (dotted line) exciton energy levels calculated with $I_{eMn}$=-0.03 meV, $I_{hMn}$=0.12 meV, $\rho_c/\Delta_{lh}=0.05$, $\eta$=10 $\mu$eV and an electron-hole exchange interaction $I_{eh}$=-0.78 meV. See for instance reference 32 for a description of the model.}
\label{FigAppendix}
\end{figure}

The mechanism of Mn-hole flip-flop that we identified and modelled in the case of positively charged magnetic QDs can also be directly observed in the resonant PL of neutral Mn-doped QDs. This is illustrated in Fig.~\ref{FigAppendix} which presents the PL of the exciton in a QD where all the dark exciton states ($|\pm2\rangle$) are well separated from the bright exciton states ($|\pm1\rangle$) and can be observed at zero magnetic field on the low energy side of the spectra.

In the PL excitation experiments presented in Fig.~\ref{FigAppendix}, the detection window is set on the dark states while a circularly cross-polarized laser is scanned on the three high energy levels of the bright exciton. Three successive resonances are observed.

(1): An excitation on $|+1,S_z=+1/2\rangle$ produces the dominant PL on a dark states associated with a Mn spin state $\pm3/2$. As a Mn spin flip by two units is unlikely, we then attribute this PL to $|-2,S_z=+3/2\rangle$. The transfer between these states involves a Mn-hole flip-flop. The second weaker contribution comes from the state $|+2,S_z=+1/2\rangle$. It corresponds to a transfer conserving the Mn spin and involving a spin flip of the electron.

(2): An excitation on $|+1,S_z=+3/2\rangle$ produces the dominant PL on a dark states associated with a Mn spin state $\pm5/2$. As discussed above, a Mn spin flip by four units is unlikely, we then attribute this PL to $|-2,S_z=+5/2\rangle$. The transfer involve a Mn-hole flip-flop. The second contribution comes from the state $|+2,S_z=+3/2\rangle$. It corresponds to a conservation of the Mn spin and a spin flip of the electron.

(3): From an excitation on $|+1,S_z=+5/2\rangle$, the dominant PL comes from the state $|+2,S_z=+5/2\rangle$. It corresponds to a transfer with conservation of the Mn spin and a spin flip of the electron. A PL is also observed from $|-2,S_z=+5/2\rangle$ after a spin flip of the hole and from $|+2,S_z=+3/2\rangle$. The latter transfer involves an electron-Mn flip-flop. Let's note that in the state $|+1,S_z=+5/2\rangle$ a Mn-hole flip-flop is forbidden (parallel hole and Mn spin).

If we resume the observed resonances in this neutral QD, we can distinguish two types of spin transfer which can occur within the lifetime of the exciton (typically 300ps): First an efficient transfer from a bright to a dark state involving a Mn-hole flip-flop which produces a change of one unit of the Mn spin state. Secondly, a weaker transfer from a bright to a dark state involving a carrier spin-flip with conservation of the Mn spin.

\end{appendix}

\end{document}